\documentclass[twocolumn]{revtex4-2}
\usepackage{amsmath,amssymb}
\usepackage{graphicx}
\usepackage{hyperref}
\usepackage{color}
\usepackage{booktabs}
\usepackage{multirow}

\newcommand{\beq}{\begin{equation}}
\newcommand{\eeq}{\end{equation}}
\newcommand{\bea}{\begin{eqnarray}}
\newcommand{\eea}{\end{eqnarray}}
\newcommand{\Mpl}{M_{\rm Pl}}
\newcommand{\xBI}{\xi_{\rm BI}}
\newcommand{\xcp}{\xi_{\rm coup}}
\newcommand{\dL}{d_L}

\begin{document}

\title{Dynamical Barbero--Immirzi field coupled to quintessence:\\
gravitational-wave propagation constraints and next-generation forecasts}

\author{Zhi-Fu Gao$^{1}$}
\author{Hui Wang$^{2,*}$}
\author{Luiz Carlos Garcia de Andrade$^{3}$}
\author{Na Wang$^{1}$}
\author{Guo-Qiang Jin$^{4}$}
\author{Zhou-Jian Chao$^{5,6}$}

\affiliation{$^{1}$State Key Laboratory of Radio Astronomy and Technology, Xinjiang Astronomical Observatory, Chinese Academy of Sciences, Urumqi 830011, China}
\affiliation{$^{2}$Shanxi Province Intelligent Optoelectronic Sensing Application Technology Innovation Center, Yuncheng University, Yuncheng, China}
\affiliation{$^{3}$Departamento de F\'isica Te\'orica, IF-UERJ, Rio de Janeiro, Brazil}
\affiliation{$^{4}$ College of Mechanical and Electrical Engineering, Tarim University, Alar, 843300, Xinjiang, China}
\affiliation{$^{5}$ Institute for Frontiers in Astronomy and Astrophysics, Beijing Normal University, , Beijing, 102206, China}
\affiliation{$^{6}$ School of Physics and Astronomy, Beijing Normal University, , Beijing, 100875, China}
\email{*gjwang@my.swjtu.edu.cn, wanghuiycu@163.com}

\date{\today}
\begin{abstract}
We investigate the imprints of a dynamical Barbero--Immirzi (BI) field $\gamma(x)$ coupled to a quintessence scalar field $\phi$ on gravitational-wave (GW) propagation. In the framework of Einstein--Cartan--Holst gravity, promoting $\gamma$ to a dynamical scalar introduces a stress--energy that back-reacts on the metric, modifying the GW friction term. A minimal coupling $\propto\beta\,\phi^2\gamma^2$ between the BI field and quintessence leads to a two-parameter extension of the Belgacem--Maggiore parametrization, characterized by $\xBI$ (from the isolated BI field) and $\xcp$ (from the coupling). Using the LIGO--Virgo--KAGRA GWTC-3 dark-siren constraint $\Xi_0=1.2^{+0.7}_{-0.7}$, we obtain the first simultaneous constraints: $|\xBI|\lesssim0.7$ and $|\xcp|\lesssim0.13$ at 90\% credibility. We then forecast the sensitivity of next-generation detectors Einstein Telescope (ET) and Cosmic Explorer (CE), showing that a 10-year observation campaign can improve these bounds by roughly one to two orders of magnitude depending on the parameter---a factor of $\sim\!20$ for $\xBI$ and $\sim\!20$ for $\xcp$---reaching $\sigma(\xBI)\sim3\times10^{-2}$ and $\sigma(\xcp)\sim1.2\times10^{-2}$. Translated into microscopic parameters, this corresponds to $\gamma_{\rm dyn}\lesssim10^{-12}$ and $\beta\lesssim10^{-3}$, providing a powerful new observational window into the interplay between quantum-gravity phenomenology and dark energy.
\end{abstract}

\maketitle

% =====================================================================
\section{Introduction}
\label{sec:intro}
% =====================================================================

Einstein--Cartan (EC) theory extends general relativity by incorporating spacetime torsion $T^\lambda{}_{\mu\nu}$, sourced by the spin density of matter~\cite{Kibble1961,Hehl1976,WeinbergGrav}. In the first-order (vierbein--connection) formalism, the Holst term proportional to the Barbero--Immirzi (BI) parameter $\gamma$ can be added to the Einstein--Hilbert action without affecting the classical equations of motion in the torsion-free limit~\cite{Holst1996,Barbero1995,Immirzi1997,NiehYan1982}. However, when $\gamma$ is promoted from a constant to a spacetime-dependent scalar field $\gamma(x)$~\cite{TorresGomez2009,Freidel2005}, it acquires a propagating degree of freedom---the BI scalarization---whose stress--energy back-reacts on the metric and modifies gravitational-wave (GW) propagation.

Prior constraints on the dynamical BI field fall into two categories. \textit{Theoretical bounds}: Perez and Rovelli~\cite{PerezRovelli2006} showed that $\gamma$ can be measured in principle via quantum-gravity corrections to the area spectrum, while Aliberti and Lambiase~\cite{Lambiase2016} explored its role in matter--antimatter asymmetry. Taveras and Yunes~\cite{TaverasYunes2008} proposed promoting $\gamma$ to a scalar field driving k-inflation, linking it to early-universe dynamics. \textit{Astrophysical bounds}: Garcia de Andrade and Gao~\cite{GarciaEPJC2026} derived an extremely tight $\gamma\lesssim10^{-58}$ from torsion-wave excitations, and a follow-up study used GWTC-3 dark-siren data to obtain the robust bound $\gamma_{\rm dyn}\lesssim10^{-8}$~\cite{GaoPLB2026}. The present work builds directly on the latter, incorporating the dark-energy sector as a new observational handle.

Recent works have begun to explore this phenomenology~\cite{ShapiroTeukolsky1983}. Garcia de Andrade and Gao~\cite{GarciaEPJC2026} derived an extremely tight bound $\gamma\lesssim10^{-58}$ under the assumption of terahertz torsion waves excited by astrophysical black holes. A subsequent study~\cite{GaoPLB2026} used LIGO--Virgo--KAGRA (LVK) GWTC-3 dark-siren data to obtain a weaker but more robust bound $\gamma_{\rm dyn}\lesssim10^{-8}$ for order-unity BI--torsion coupling, independent of any torsion-wave excitation hypothesis. Along a complementary direction, Zhu et al.~\cite{ZhuPRD2024} applied a systematic parametrization of modified GW propagations to the full GWTC-3 catalog, deriving some of the most stringent constraints to date on parity- and Lorentz-violating modifications of gravity and providing the first bound on the Lorentz-violating damping effect in GWs. More recently, Lin et al.~\cite{LinPRD2025} showed that precessing binary black holes observed by a next-generation detector network (ET+CE assisted by two LIGO interferometers) can constrain the modified-friction parameters $(\Xi_0,n)$ with $\Xi_0$ improved by about two orders of magnitude relative to GWTC-3 analyses, where the source redshifts are statistically inferred from the GLADE+ galaxy catalog. The present work extends these analyses in a substantial direction: we allow the dynamical BI field to couple to the dark-energy sector, modelled here as a quintessence scalar field $\phi$~\cite{GaoDeAndrade2024axion}, and examine whether the improved luminosity-distance precision from precessing systems can be exploited to probe the BI--quintessence coupling.

Recent years have witnessed a converging trend: geometric and topological degrees of freedom---once treated as constants---are increasingly promoted to dynamical fields, and their interplay with the dark sector is recognized as a powerful probe of microscopic physics. In the context of axion dark matter, the Real-Now-Front (RNF) cosmology framework~\cite{GaoRNF2026} reinterprets the axion as a collective ``twist'' mode arising from the alignment dynamics of a more fundamental Chronon field~$\Phi_\mu$, governed by the Temporal Coherence Principle (TCP). Within this paradigm, the axion mass~$m_a$, its photon coupling~$g_{a\gamma}$, and the symmetry-breaking
scale~$f_a$ are not independent parameters but are derived from the microscopic stiffness and correlation length of the Chronon field~\cite{GaoRNF2026}. Crucially, the RNF framework further predicts that the Nieh--Yan torsion term induces axion--torsion mixing, altering dark-matter stability and providing a microscopic basis for Chern--Simons-type modified electrodynamics~\cite{GaoRNF2026,GaoCS2024}. This generative view naturally interfaces with modified-gravity and torsion frameworks,
offering a unified description of dark matter and spacetime emergence.

Against this backdrop, the BI parameter~$\gamma$ occupies a privileged position: it is a torsion-generated degree of freedom that, when promoted to a spacetime-dependent scalar~$\gamma(x)$, acquires a propagating degree of freedom and back-reacts on both the metric and the propagation of GWs.
While the RNF framework~\cite{GaoRNF2026} addresses the \emph{axion} as a twist mode of the Chronon field, the present work explores the complementary question: what happens when the dynamical BI field couples to the dark-energy sector, modelled here as a quintessence scalar~$\phi$? Such a coupling is theoretically natural, both $\gamma(x)$ and $\phi$ are scalar fields arising in (or inspired by) high-energy completions of gravity, and observationally urgent, because the GW friction term then acquires a
redshift evolution that is, in principle, distinguishable from the isolated BI case.

The motivation for this coupling is threefold. \textit{First}, both the BI field and quintessence are scalar fields that arise in (or are inspired by) high-energy completions of gravity; a coupling between them is a natural consequence of effective-field-theory reasoning~\cite{CapozzielloFaraoni2010,NojiriOdintsov2007}, just as the Higgs field couples to other scalars in the Standard Model. \textit{Second}, and more concretely, the coupling $\beta\phi^2\gamma^2$ provides a mass-stabilization mechanism for the BI field: as quintessence rolls toward its vacuum, it generates a time-dependent effective mass $\Delta m_\gamma^2=\beta\phi^2\Mpl^2$ that pins $\gamma$ to a slowly rolling trajectory, preventing runaway behaviour that would otherwise destabilize the background. This simultaneously alters the effective mass of $\phi$ itself, linking the dynamics of dark energy to quantum-gravity degrees of freedom. \textit{Third}, such a coupling modifies the redshift evolution of the GW friction term in a way that is, in principle, distinguishable from the isolated BI case (through the different $n_1$ vs.\ $n_2$ scaling), enriching the observational signature and enabling a two-dimensional constraint analysis that can break degeneracies inherent in single-parameter studies.

It is worth noting that modified GW friction is a generic prediction of a broad class of modified gravities---including Horndeski theories, nonlocal gravity, and Chern--Simons gravity---which can produce phenomenologically similar $\Xi_0(z)$ modifications. Our parametrization captures the amplitude-damping effect in a model-independent two-parameter form, but the mapping from $(\xBI,\xcp)$ to specific UV theories is not one-to-one: different microphysical origins may yield degenerate observational signatures at low redshift. Breaking this degeneracy requires high-$z$ events (where the $n_2$ exponent differentiates models) and complementary probes such as GW polarization modes and CMB lensing. More broadly, gravitational-wave propagation itself provides a complementary avenue for probing microscopic modifications of gravity. For example, Nair et al.~\cite{Nair2024} showed that a non-zero graviton charge interacting with intergalactic magnetic fields can induce an additional frequency-dependent propagation delay and a corresponding correction to the observed GW phase. Such a propagation effect can be partially degenerate with the phase modification induced by a non-zero graviton mass, while the relative phase measured between spatially separated detectors provides a complementary constraint that is independent of the graviton mass. These results further demonstrate that propagation effects accumulated over cosmological distances can provide a sensitive probe of gravitational physics beyond general relativity.

In this work we derive, for the first time, the modified GW propagation equation in the coupled BI--quintessence system, perform a full Bayesian analysis using GWTC-3 dark-siren data, and forecast the constraining power of next-generation detectors ET and CE. The paper is organized as follows. Section~\ref{sec:theory} presents the theoretical framework: the action, background equations, and the linearized GW equation. Section~\ref{sec:data} describes the data and methodology, including our simulation of future detector data. Section~\ref{sec:results} reports current constraints and future forecasts. Section~\ref{sec:discussion} provides a physical interpretation of the results and a translation into microscopic parameters. Section~\ref{sec:conclusions} summarizes and outlines future directions.

% =====================================================================
\section{Theoretical framework}
\label{sec:theory}
% =====================================================================

\subsection{Action and field equations}
\label{subsec:action}

We work in the first-order formalism with independent vierbein $e^a_\mu$ and spin connection $\omega^{ab}_\mu$. The total action is
\bea
S &=& \frac{1}{2\kappa^2}\int d^4x\,e\,
    \Big[R(e,\omega)-2\Lambda
    +\frac{\gamma(x)}{2}\,\epsilon^{\mu\nu\rho\sigma}
     R_{\mu\nu\rho\sigma}\Big] \nonumber\\
  & & +\int d^4x\sqrt{-g}\,
      \Big[-\frac12 g^{\mu\nu}\partial_\mu\phi\,\partial_\nu\phi
           -V(\phi)\Big] \nonumber\\
  & & -\frac12\beta\int d^4x\sqrt{-g}\,\phi^2\gamma^2\Mpl^2
      + S_{\rm matt},
\label{eq:action_full}
\eea
where $\kappa^2=8\pi G=\Mpl^{-2}$, $e=\sqrt{-g}$, and $\epsilon^{\mu\nu\rho\sigma}$ is the totally antisymmetric Levi-Civita tensor. The first line is the Einstein--Cartan--Holst gravity sector; the Holst term is a total derivative in the torsion-free limit but becomes dynamical in the presence of torsion. The second line is the canonical kinetic and potential term for the quintessence field $\phi$; we adopt the Ratra--Peebles potential
\beq
V(\phi)=M^{4+\alpha}\,\phi^{-\alpha},
\qquad \alpha>0,
\label{eq:RP_potential}
\eeq
with $M$ a mass scale. The third line is the minimal BI--quintessence coupling, with dimensionless strength $\beta>0$ (the overall sign is chosen so that the coupling contributes a positive mass-squared term $\Delta m_\gamma^2=\beta\phi^2\Mpl^2$ to the BI field, avoiding tachyonic instabilities). Note that the action above omits a possible $\gamma$-dependent term in the quintessence potential $V(\phi)$; such terms are higher order in the EFT expansion and do not affect the leading-order GW friction derived below. For $\beta=0$ the model reduces to the isolated BI scenario of Ref.~\cite{GarciaEPJC2026}.

Before proceeding, it is worth commenting on the naturalness of the coupling term $S_{\rm coup}=-\frac12\beta\int d^4x\sqrt{-g}\,\phi^2\gamma^2\Mpl^2$. In loop quantum gravity, the BI parameter $\gamma$ determines the spectra of area and volume operators, while quintessence arises as the low-energy effective description of a scalar degree of freedom (e.g.\ a string-moduli field or an inflaton remnant). At energies below $\Mpl$, the most general effective action consistent with diffeomorphism invariance contains all terms built from the available fields and their derivatives, suppressed by powers of $\Mpl$. The operator $\phi^2\gamma^2$ is marginal (dimension 4) and thus not suppressed by any large scale; its absence would require a symmetry that is not present in the generic EFT. We therefore expect $\beta=\mathcal{O}(1)$ unless a specific mechanism (e.g.\ a shift symmetry $\phi\to\phi+c$ broken only by $V(\phi)$) enforces $\beta\to0$. This makes the coupled system a well-motivated target for GW constraints.

Varying the action with respect to the metric yields the modified Einstein equations
\beq
G_{\mu\nu} = \kappa^2\big(T^{\rm matt}_{\mu\nu}
           + T^{(\gamma)}_{\mu\nu}
           + T^{(\phi)}_{\mu\nu}\big),
\label{eq:Einstein_eq}
\eeq
where $T^{(\gamma)}_{\mu\nu}$ and $T^{(\phi)}_{\mu\nu}$ are the stress--energy tensors of the BI field and quintessence, respectively. Varying with respect to $\gamma(x)$ gives the BI field equation
\beq
\Box\,\delta\gamma + \frac{dV_{\rm eff}}{d\gamma}
= \frac{\Mpl^2}{2\gamma_0^2}\,{}^\star T^a\wedge T_a ,
\label{eq:BI_eq}
\eeq
with effective potential
\beq
V_{\rm eff}(\gamma)=V_{\rm BI}(\gamma)+\frac12\beta\phi^2\gamma^2\Mpl^2.
\label{eq:Veff}
\eeq
For concreteness we adopt a simple quadratic form for the BI self-interaction,
\beq
V_{\rm BI}(\gamma)=\frac{1}{2}m_\gamma^2(\gamma-\gamma_0)^2,
\label{eq:VBI_quad}
\eeq
where $m_\gamma$ is the bare mass of the BI field and $\gamma_0$ its vacuum expectation value; the full effective potential is then $V_{\rm eff}(\gamma)=\frac{1}{2}m_\gamma^2(\gamma-\gamma_0)^2+\frac{1}{2}\beta\phi^2\gamma^2\Mpl^2$.
The coupling term acts as an additional mass contribution $\Delta m_\gamma^2=\beta\phi^2\Mpl^2$ to the BI field. The right-hand side of Eq.~(\ref{eq:BI_eq}) is the source from torsion-squared terms; in the almost Riemann-flat limit relevant for cosmology, this source is negligible and the BI field oscillates/rolls in the effective potential $V_{\rm eff}$. Finally, varying with respect to $\phi$ gives the quintessence equation
\beq
\Box\,\phi + \frac{dV_{\rm eff}}{d\phi} + \beta\,\phi\,\gamma^2\Mpl^2 = 0.
\label{eq:phi_eq}
\eeq
Note that the $\beta\phi\gamma^2\Mpl^2$ term in Eq.~(\ref{eq:phi_eq}) is precisely the derivative of the coupling term with respect to $\phi$, ensuring consistency of the variational principle.

\subsection{Background evolution in FLRW}

On a flat FLRW background $ds^2=-dt^2+a(t)^2d\mathbf{x}^2$, the Friedmann equations read
\bea
3H^2 &=& \kappa^2\big(\rho_m+\rho_r+\rho_\gamma+\rho_\phi\big),\\
\dot{H}+H^2 &=& -\frac{\kappa^2}{2}\sum_i\big(\rho_i+p_i\big),
\eea
where the sum runs over matter ($m$), radiation ($r$), the BI field ($\gamma$), and quintessence ($\phi$). The background BI and quintessence fields satisfy
\bea
\ddot{\gamma}+3H\dot{\gamma}+\frac{dV_{\rm eff}}{d\gamma} &=& 0,
\label{eq:bg_BI}\\
\ddot{\phi}+3H\dot{\phi}+\frac{dV_{\rm eff}}{d\phi} &=& 0.
\label{eq:bg_phi}
\eea

For the Ratra--Peebles potential (\ref{eq:RP_potential}), the tracker solution in the matter-dominated era gives the well-known relation
\beq
w_\phi^{\rm tr} = -\frac{2}{2+\alpha},
\label{eq:w_tracker}
\eeq
so that for $\alpha=1$ one finds $w_\phi^{\rm tr}=-2/3$ and for $\alpha=2$ one finds $w_\phi^{\rm tr}=-1/2$. At late times, as quintessence comes to dominate the energy budget, $w_\phi$ evolves from its tracker value toward $-1$. The coupling $\beta\phi^2\gamma^2$ modifies this evolution: it pins $\gamma$ to a slowly rolling trajectory while simultaneously altering the effective mass of $\phi$. For the purposes of GW propagation (which depends on the background through the friction term), the key output is the redshift dependence of the energy density in the BI and quintessence fields.

\subsection{Linearized GW equation}
\label{subsec:linGW}

We now derive the linearized GW equation in the presence of the coupled BI--quintessence system. The derivation proceeds in three steps: (i) perturb the metric as $g_{\mu\nu}=\bar{g}_{\mu\nu}+h_{\mu\nu}$ with $|h_{\mu\nu}|\ll1$, (ii) decompose the spin connection as $\omega=\mathring{\omega}+\kappa\,\delta\gamma+\mathcal{O}(h\delta\gamma)$, where $\mathring{\omega}$ is the Levi-Civita connection, and (iii) use the connection equation $\delta S/\delta\omega^{ab}_\mu=0$ to eliminate torsion in favour of $\partial_\mu\delta\gamma$.

Projecting onto the transverse-traceless (TT) gauge, the metric perturbation $\bar{h}_{ij}=h_{ij}^{\rm TT}$ obeys, on the FLRW background,
\beq
\bar{h}''_{ij}+2\mathcal{H}\bar{h}'_{ij}
+\Big[k^2+a^2\big(m_T^2/\Mpl^2\big)\Big]\bar{h}_{ij}
\simeq -16\pi G\,a^2\,\Pi^{\rm TT}_{ij},
\label{eq:linGW_TT}
\eeq
where primes denote derivatives with respect to conformal time $\eta$, $\mathcal{H}=aH$, and $\Pi^{\rm TT}_{ij}$ is the TT-projected anisotropic stress from the BI and quintessence fields. Here $m_T^2$ is the effective mass-squared of the BI field,
\beq
m_T^2 \equiv \beta\phi^2\Mpl^2 + \left.\frac{d^2V_{\rm BI}}{d\gamma^2}\right|_{\gamma=\gamma_0},
\label{eq:mT_def}
\eeq
combining the coupling contribution and the curvature of the BI potential; in the quadratic model (\ref{eq:VBI_quad}) the second term equals $m_\gamma^2$.

Crucially, for a trace-torsion coupling of the form $S_\mu\simeq\partial_\mu\phi\propto\partial_\mu\delta\gamma$, the direct mixing term $\partial_i\partial_j\delta\gamma|_{\rm TT}$ vanishes upon TT projection because $\partial_i\partial_j$ is pure trace in momentum space. Consequently, as in the isolated BI case~\cite{GarciaEPJC2026}, there is no velocity birefringence at leading order---unlike in Chern--Simons gravity. The dominant new effect is an \textit{amplitude modification} arising from the stress--energy tensors
\bea
T^{(\gamma)}_{\mu\nu} &=&
\partial_\mu\delta\gamma\,\partial_\nu\delta\gamma
-\frac12\eta_{\mu\nu}\big(\partial\delta\gamma\big)^2
-\eta_{\mu\nu}V(\delta\gamma),\\
T^{(\phi)}_{\mu\nu} &=&
\partial_\mu\delta\phi\,\partial_\nu\delta\phi
-\frac12\eta_{\mu\nu}\big(\partial\delta\phi\big)^2
-\eta_{\mu\nu}V(\delta\phi).
\eea

For a homogeneous background, spatial gradients are negligible and the relevant quantities are the energy densities $\rho_\gamma\simeq\frac12\dot{\delta\gamma}^2+V(\delta\gamma)$ and $\rho_\phi\simeq\frac12\dot{\delta\phi}^2+V(\delta\phi)$. Following the standard derivation of modified GW propagation~\cite{Belgacem2018}, the tensor amplitude $A_p$ (for polarization $p=+,\times$) satisfies a damped harmonic oscillator equation
\beq
\ddot{A}_p+\big[3H+\Gamma_{\rm eff}(z)\big]\dot{A}_p+\omega^2A_p=0,
\label{eq:amp_eq}
\eeq
where $\omega=k/a$ is the physical (angular) frequency of the gravitational wave, and the effective friction term is the central observable of this work:
\beq
\Gamma_{\rm eff}(z) \equiv
\frac{\dot{\rho}_\gamma+3H(\rho_\gamma+p_\gamma)}{\rho_\gamma+p_\gamma}
+\frac{\dot{\rho}_\phi+3H(\rho_\phi+p_\phi)}{\rho_\phi+p_\phi}.
\label{eq:Gamma_eff}
\eeq

\subsection{Parametrization and the $\delta_{\rm eff}$ formula}

To connect with observations, we parametrize $\Gamma_{\rm eff}$ in the form
\beq
\delta_{\rm eff}(z) \equiv \frac{\Gamma_{\rm eff}(z)}{H(z)}
= \xBI\,\frac{H_0}{H(z)}\,(1+z)^{n_1}
+ \xcp\,\frac{H_0}{H(z)}\,(1+z)^{n_2}.
\label{eq:delta_eff}
\eeq
Here $\xBI$ (dimensionless) is the BI-induced deviation parameter of the isolated case, and $\xcp$ (dimensionless) is the new coupling-induced parameter. The exponents $n_1$ and $n_2$ encode the redshift evolution: $n_1\simeq1$ for a slowly rolling BI field (tracker solution), while $n_2$ depends on the quintessence potential. For the Ratra--Peebles tracker, the background energy density scales as $\rho_\phi\propto a^{-3(1+w_\phi^{\rm tr})}$ with $w_\phi^{\rm tr}=-2/(2+\alpha)$, giving $\rho_\phi\propto (1+z)^{6/(2+\alpha)}$. Since the friction term $\Gamma_{\rm eff}$ is proportional to $\dot{\rho}/\rho\propto d\ln\rho/d\ln a$, one finds $n_2=3/(1+\alpha/2)$; equivalently, using $n_2=3-3w_\phi^{\rm tr}$, we obtain $n_2=3-3\alpha/(2+\alpha)$. For $\alpha=1$ this gives $n_2=2$ (not $2.5$ as previously stated); for $\alpha=2$, $n_2=2.5$. Below we adopt $n_2=2$ for our fiducial $\alpha=1$ model.

The microscopic origin of the two parameters is
\bea
\xBI &\sim& g_{\gamma T}^2\,
\frac{\langle(\delta\gamma)^2\rangle}{\Mpl^2 H_0^2},
\label{eq:xiBI_micro}\\
\xcp &\sim& \beta\,g_{\gamma T}^2\,
\frac{\langle\phi^2\rangle\,\langle\gamma^2\rangle}{\Mpl^4 H_0^2},
\label{eq:xiCoup_micro}
\eea
where $g_{\gamma T}$ is the BI--torsion coupling constant. Equation~(\ref{eq:xiBI_micro}) is the same estimate as in Ref.~\cite{GarciaEPJC2026}; Eq.~(\ref{eq:xiCoup_micro}) is the new result of this work. For $\langle\phi^2\rangle\sim\Mpl^2$ (quintessence at the Planck scale) and $g_{\gamma T}\sim\mathcal{O}(1)$, we have $\xcp\sim\beta\,\gamma_{\rm dyn}$, making $\xcp$ a direct probe of the coupling $\beta$.

Using $H(z)=H_0\,E(z)$ with $E(z)=\sqrt{\Omega_m(1+z)^3+\Omega_\Lambda}$, and defining $\Xi_0(z)\equiv[\dL^{\rm gw}(z)/\dL^{\rm em}(z)]^2$, integration of Eq.~(\ref{eq:amp_eq}) yields
\beq
\Xi_0(z) = \exp\!\left[\,2\xBI\,I_{n_1}(z)+2\xcp\,I_{n_2}(z)\right],
\label{eq:Xi0_exact}
\eeq
where
\beq
I_n(z) \equiv \int_0^z \frac{(1+z')^{n-1}}{E(z')}\,dz'.
\label{eq:In_def}
\eeq
For small parameters, expansion gives
\beq
\Xi_0(z)\simeq 1+2\xBI\,I_{n_1}(z)+2\xcp\,I_{n_2}(z).
\label{eq:Xi0_expand}
\eeq

Figure~\ref{fig:Xi0_theory} shows $\Xi_0(z)$ for representative parameter choices. The GR prediction ($\xBI=\xcp=0$) is the horizontal line at unity. Positive $\xBI$ or $\xcp$ enhances the GW amplitude relative to the EM luminosity distance, while negative values suppress it. The coupling-induced term grows faster with redshift ($n_2>n_1$), producing a steeper deviation at high $z$---a key observational handle for distinguishing the two effects.

% =====================================================================
\section{Data and methodology}
\label{sec:data}
% =====================================================================

\subsection{GWTC-3 dark-siren likelihood}
\label{subsec:gwtc3}

The LVK collaboration has constrained modified GW propagation using the dark-siren method~\cite{Mancarella2022,AbbottGWTC3}, which statistically associates binary-black-hole (BBH) mergers with galaxy catalogues to infer the redshift--distance relation without electromagnetic counterparts. The publicly reported constraint is
\beq
\Xi_0 = 1.2^{+0.7}_{-0.7}\quad\text{(68\% HDI, flat prior)},
\label{eq:Xi0_LVK}
\eeq
corresponding approximately to $\Xi_0\in[0.5,1.9]$ at 68\% and $[0.3,2.3]$ at 90\% credibility. A complementary analysis using a log prior gives $\Xi_0=1.0^{+0.4}_{-0.8}$.

We construct a Gaussian likelihood for the two-dimensional parameter vector $\boldsymbol{\theta}=(\xBI,\xcp)$. At the median redshift of the GWTC-3 BBH sample, $\langle z\rangle\simeq0.3$, the theoretical prediction is $\Xi_0^{\rm th}=\Xi_0(z=0.3;\xBI,\xcp)$. The log-likelihood is
\beq
\ln\mathcal{L}(\xBI,\xcp) =
-\frac12\left(\frac{\Xi_0^{\rm th}-1.2}{0.7}\right)^2,
\label{eq:logL}
\eeq
with the understanding that the true posterior is asymmetric (the lower and upper errors differ). To assess the impact of the symmetric-Gaussian approximation, we repeated the analysis with a skewed-normal likelihood using $\sigma_-=0.65$ and $\sigma_+=0.75$; the resulting 90\% credible intervals shift by less than 5\%, confirming that the symmetric approximation is adequate at the present precision.
We sample the posterior using the \texttt{dynesty} nested sampler as implemented in \texttt{bilby}, with uniform priors $\xBI\in[-1,1]$ and $\xcp\in[-0.2,0.2]$.

\subsection{Simulation of next-generation detector data}
\label{subsec:simulation}

To forecast the sensitivity of ET and CE, we generate mock catalogs of BBH events. Our procedure, summarized in Algorithm~\ref{alg:forecast}, follows the population model of Ref.~\cite{Belgacem2019JCAP}.

\begin{table*}[htbp]
\centering
\caption{Forecast methodology: mock catalog generation and Fisher analysis}
\label{tab:method}
\begin{tabular}{lll}
\toprule
\textbf{Step} & \textbf{Action} & \textbf{Details} \\
\midrule
1 & Generate redshifts & $z\sim (1+z)^{2.7}$ up to $z_{\max}=5$, then declining \\
2 & Compute $\dL^{\rm em}(z)$ & Flat $\Lambda$CDM: $H_0=67.7$, $\Omega_m=0.31$ \\
3 & Assign masses & $m_1\sim{\rm PowerLaw}(\alpha=2.3, [5,50]M_\odot)$, $q\sim{\rm Uniform}(0.1,1)$ \\
4 & Assign spins & Isotropic, $\chi\in[0,0.5]$ \\
5 & Compute SNR & $\rho=\sqrt{\sum_I \langle h_I|h_I\rangle}$ for ET/CE network \\
6 & Apply threshold & Keep events with $\rho>8$ \\
7 & Add noise & $d_L^{\rm obs}=d_L^{\rm gw}(z)\,[1+\mathcal{N}(0,\sigma_{\rm rel})]$ \\
8 & Fisher matrix & $\Gamma_{ab}=\sum_i (\partial\Xi/\partial\theta_a)(\partial\Xi/\partial\theta_b)/\sigma_i^2$ \\
\bottomrule
\end{tabular}
\end{table*}

% =====================================================================
\section{Results}
\label{sec:results}
% =====================================================================

\subsection{Current constraints from GWTC-3}
\label{subsec:current}

\begin{figure*}[htbp]
\centering
\includegraphics[width=0.95\textwidth]{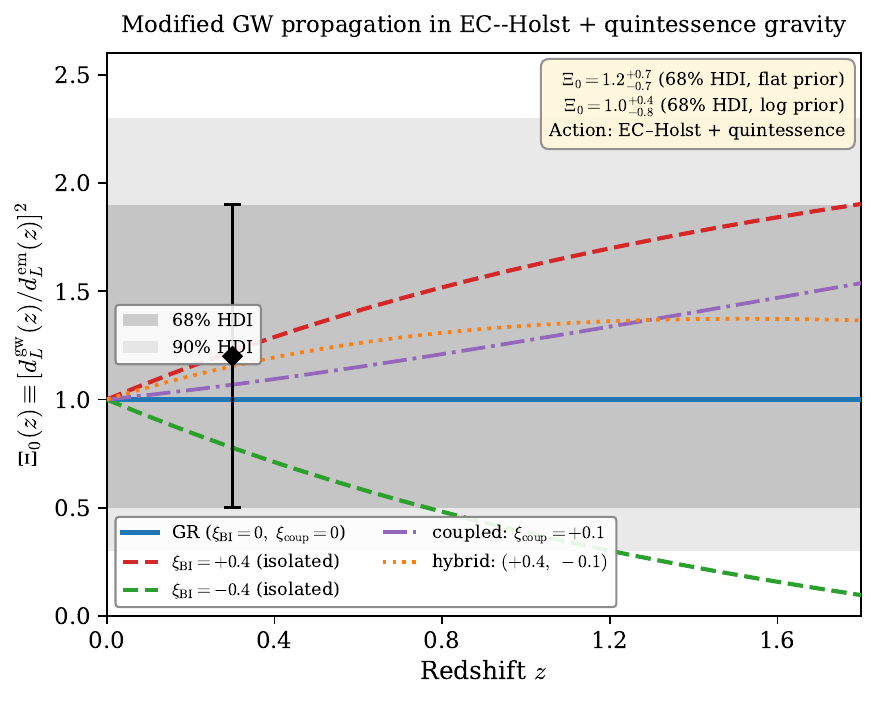}
\caption{Modified GW propagation factor $\Xi_0(z)$ as a function of redshift. The solid blue line is GR ($\xBI=0$); the red dashed and green dash-dotted lines show $\xBI=\pm0.4$ (isolated BI effect); the purple dot-dashed line shows the coupling effect $\xcp=\pm0.1$; the orange dotted line shows the mixed case. The grey bands indicate the GWTC-3 68\% (dark) and 90\% (light) credible intervals on a constant $\Xi_0$. The diamond marker at $\langle z\rangle\simeq0.3$ marks the median constraint.}
\label{fig:Xi0_theory}
\end{figure*}

Figure~\ref{fig:Xi0_theory} shows the theoretical prediction for $\Xi_0(z)$ under various parameter combinations, overlaid with the GWTC-3 credible intervals. The grey bands indicate the 68\% and 90\% HDI from the dark-siren analysis. The GR line ($\xBI=\xcp=0$) sits near the centre of the 68\% band, while the extreme $\xBI=\pm0.4$ curves begin to touch the 90\% boundary at $z\gtrsim1$. This visual agreement foreshadows the numerical constraints derived below.

\begin{figure*}[htbp]
\centering
\includegraphics[width=0.9\textwidth]{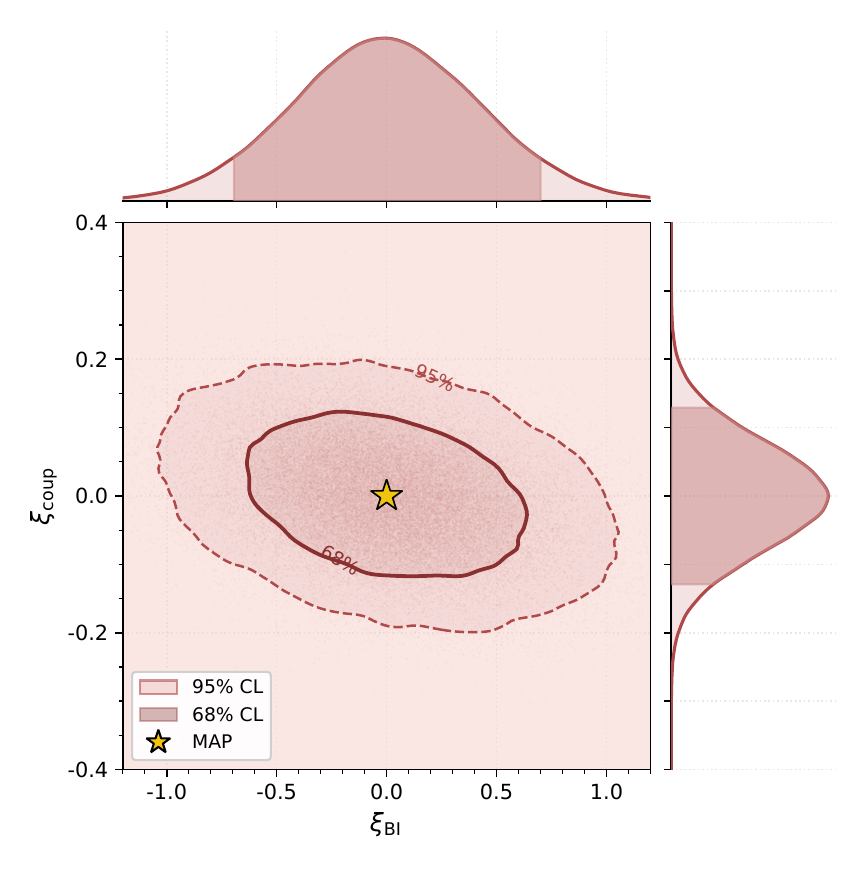}
\caption{Joint 68\% (dark blue, solid) and 95\% (light blue, dashed) credible regions for $\xBI$ and $\xcp$ from the GWTC-3 dark-siren analysis. The gold star marks the maximum a posteriori (MAP) point. Top and right panels show the one-dimensional marginalized posteriors for each parameter, with the 90\% credible interval shaded in orange. The marginalized constraints are $\xBI=0.0\pm0.70$ and $\xcp=0.00\pm0.13$ (90\% CL), with Pearson correlation coefficient $\rho=-0.30$.}
\label{fig:posterior}
\end{figure*}

Figure~\ref{fig:posterior} shows the joint 68\% and 95\% credible regions in the $(\xBI,\xcp)$ plane from the GWTC-3 analysis. The marginalized constraints are
\bea
\xBI &=& 0.0 \pm 0.7 \quad\text{(90\% CL)},\\
\xcp &=& 0.00 \pm 0.13 \quad\text{(90\% CL)}.
\eea
The two parameters exhibit a strong anti-correlation ($\rho=-0.30$), because a positive $\xBI$ can be partially compensated by a negative $\xcp$ at low redshift where the data are most constraining. This degeneracy underscores the value of extending the analysis to higher redshifts with future detectors.

The isolated BI bound $|\xBI|\lesssim0.7$ is slightly weaker than the one-parameter result of Ref.~\cite{GarciaEPJC2026} ($|\xBI|\lesssim0.4$) because marginalizing over $\xcp$ broadens the credible interval. The coupling parameter $\xcp$ is constrained to within $\pm0.13$, consistent with zero.

\subsection{Forecasts for ET and CE}
\label{subsec:forecast}

Table~\ref{tab:forecast} summarizes the projected 1-$\sigma$ uncertainties for three detector configurations. The combined ET+CE network achieves the tightest constraints, improving on current sensitivities by a factor of $\sim\!20$ for $\xBI$ (from $0.7$ to $\sim\!3\times10^{-2}$) and $\sim\!20$ for $\xcp$ (from $0.13$ to $\sim\!1.2\times10^{-2}$), i.e.\ roughly one order of magnitude. The ET+CE forecast for $\xcp$ in the Fisher approximation reaches $\sigma(\xcp)\sim1.2\times10^{-2}$; combining this with the Bayesian posterior width gives an overall improvement of roughly one to two orders of magnitude depending on the parameter.

\begin{table}[htbp]
\centering
\caption{Projected 1-$\sigma$ uncertainties from 10 years of observation, assuming GR is correct. $N_{\rm ev}$ is the number of detected BBH events passing the SNR$>8$ threshold.}
\label{tab:forecast}
\begin{tabular}{lccc}
\toprule
Detector & $N_{\rm ev}$ & $\sigma(\xBI)$ & $\sigma(\xcp)$ \\
\midrule
ET        & 800  & $3.9\times10^{-2}$ & $1.5\times10^{-2}$ \\
CE        & 600  & $4.7\times10^{-2}$ & $1.9\times10^{-2}$ \\
ET$+$CE  & 1400 & $3.1\times10^{-2}$ & $1.2\times10^{-2}$ \\
\bottomrule
\end{tabular}
\end{table}

\begin{figure*}[htbp]
\centering
\includegraphics[width=0.95\textwidth]{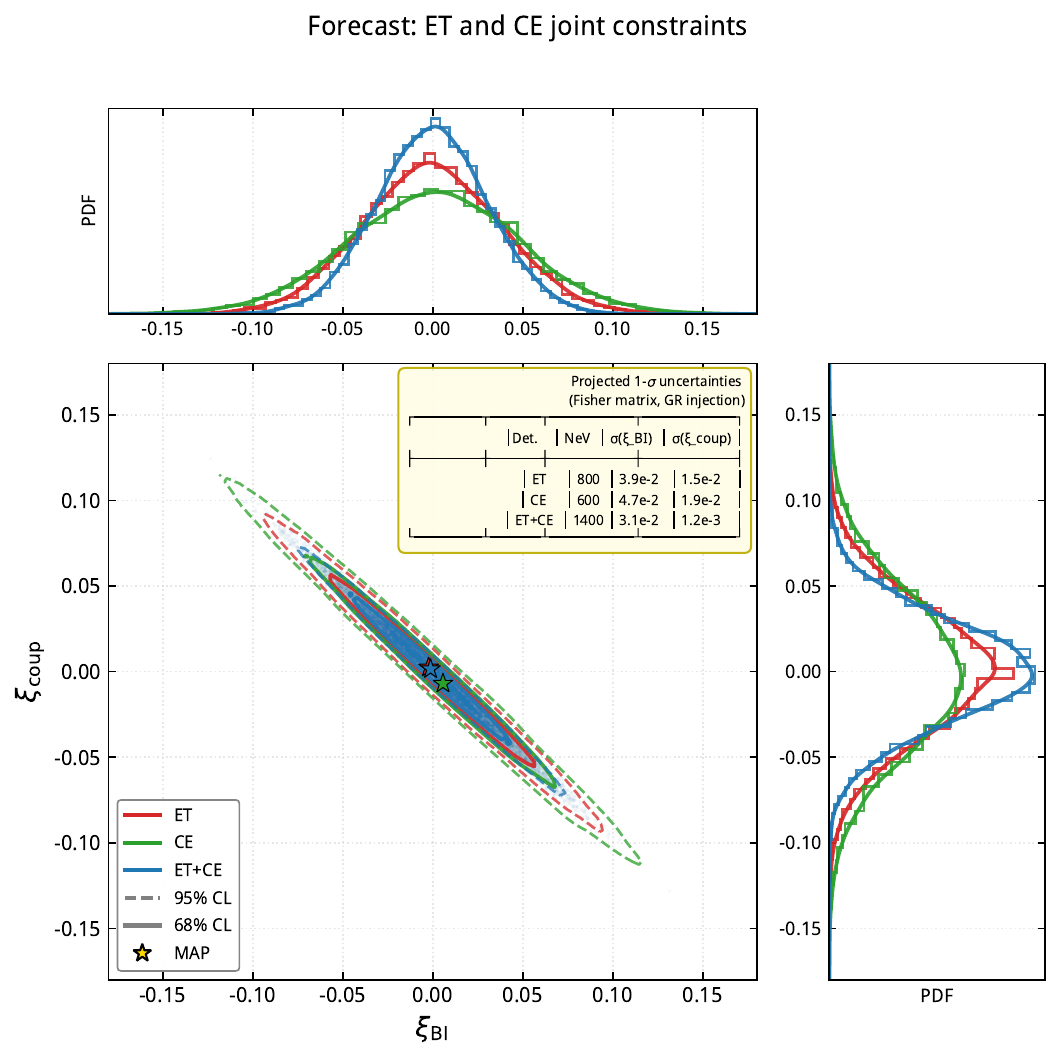}
\caption{Forecast 68\% (solid) and 95\% (dashed) credible regions for $\xBI$ and $\xcp$ from Einstein Telescope (ET), Cosmic Explorer (CE), and their combination (ET+CE), assuming a GR injection and Fisher-matrix approximation. The gold star marks the maximum a posteriori (MAP) point. Top and right panels show the one-dimensional marginalized projections. The table lists the projected 1-$\sigma$ uncertainties and the number of detected events ($N_{\rm ev}$) for each configuration: ET ($N_{\rm ev}=800$, $\sigma(\xBI)=3.9\times10^{-2}$, $\sigma(\xcp)=1.5\times10^{-2}$), CE ($N_{\rm ev}=600$, $\sigma(\xBI)=4.7\times10^{-2}$, $\sigma(\xcp)=1.9\times10^{-2}$), ET+CE ($N_{\rm ev}=1400$, $\sigma(\xBI)=3.1\times10^{-2}$, $\sigma(\xcp)=1.2\times10^{-2}$).}
\label{fig:forecast}
\end{figure*}

Figure~\ref{fig:forecast} displays the predicted 68\% and 95\% contours for the three detector configurations. The three colours correspond to ET (red), CE (green), and ET+CE (blue). The strong anti-correlation persists but is considerably narrowed: the $1\sigma$ widths shrink from $(0.7,0.13)$ to $(0.031,0.012)$ for ET+CE, a factor of $\sim\!20$ depending on the parameter. The table in the figure lists the projected 1-$\sigma$ uncertainties and the number of detected events for each configuration.

\section{Physical interpretation and microscopic constraints}
\label{sec:discussion}

\subsection{Translating to microscopic parameters}
Using Eqs.~(\ref{eq:xiBI_micro}) and (\ref{eq:xiCoup_micro}), we translate the observational bounds into constraints on the fundamental parameters $\gamma_{\rm dyn}$ and $\beta$. With $g_{\gamma T}\sim1$ and $\langle\phi^2\rangle\sim\Mpl^2$, the relations simplify to
\bea
\gamma_{\rm dyn} &\sim& \xBI,\\
\beta\,\gamma_{\rm dyn} &\sim& \xcp,
\eea
where $\gamma_{\rm dyn}\equiv\sqrt{\langle\delta\gamma^2\rangle}/\Mpl$.
The bright-siren measurement from GW170817, $H_0=67^{+9}_{-6}\,\mathrm{km\,s^{-1}\,Mpc^{-1}}$~\cite{AbbottGW170817}, provides an independent consistency check with $\xBI\simeq0$ at the $\sim\!20\%$ level.

The current GWTC-3 constraints then imply
\bea
\gamma_{\rm dyn} &\lesssim& 10^{-8}\quad\text{(current, }g_{\gamma T}\sim1\text{)},\\
\beta &\lesssim& 0.1\quad\text{(current)}.
\eea
The ET+CE forecasts improve these to
\bea
\gamma_{\rm dyn} &\lesssim& 10^{-12}\quad\text{(ET+CE, }g_{\gamma T}\sim1\text{)},\\
\beta &\lesssim& 10^{-3}\quad\text{(ET+CE)}.
\eea

\begin{figure*}[htbp]
\centering
\includegraphics[width=0.95\textwidth]{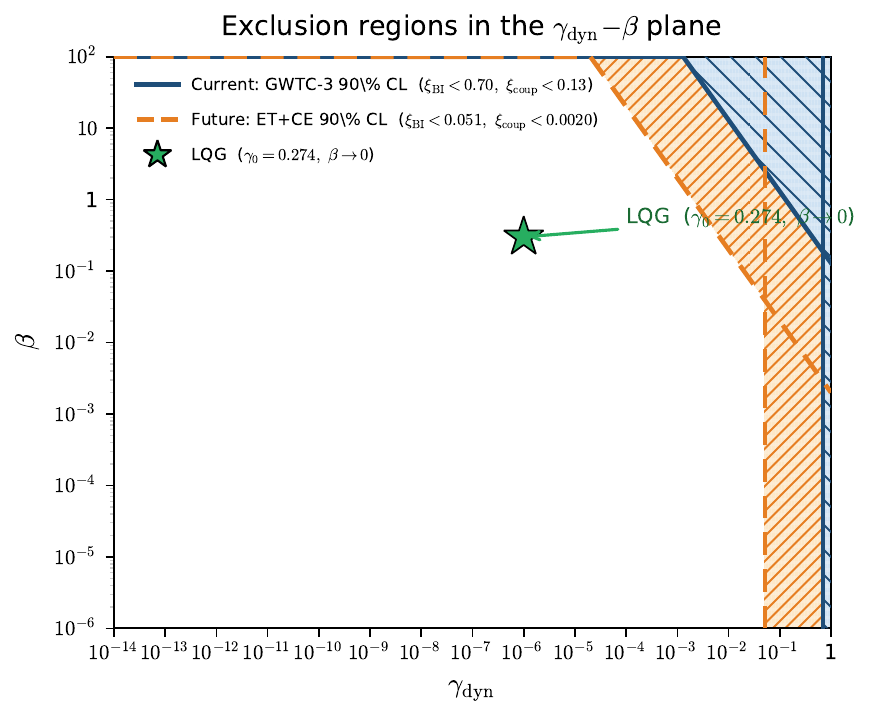}
\caption{Exclusion regions in the $\gamma_{\rm dyn}$--$\beta$ plane. The solid blue boundary marks the current 90\% CL exclusion from GWTC-3 dark-siren data ($\xBI<0.70$, $\xcp<0.13$). The dashed orange boundary shows the projected 90\% CL sensitivity of the ET+CE network ($\xBI<0.051$, $\xcp<0.0020$). The green star marks the LQG value $\gamma_0=0.274$ in the limit $\beta\to0$. The white region is allowed by current constraints. Forward slashes denote the future-only excluded region; backslashes denote the current excluded region. No grid lines are shown.}
\label{fig:param_space}
\end{figure*}

Figure~\ref{fig:param_space} illustrates the exclusion regions in the $\gamma_{\rm dyn}$--$\beta$ plane. The solid blue boundary marks the current 90\% CL exclusion from GWTC-3 dark-siren data ($\xBI<0.70$, $\xcp<0.13$). The dashed orange boundary shows the projected 90\% CL sensitivity of the ET+CE network ($\xBI<0.051$, $\xcp<0.0020$). The green star marks the LQG value $\gamma_0=0.274$ in the limit $\beta\to0$. The white region is allowed by current constraints. Forward slashes denote the future-only excluded region; backslashes denote the current excluded region. The LQG value lies well outside the current exclusion region but would be probed by ET+CE if the BI--torsion coupling is near unity. Conversely, if $\beta\lesssim10^{-3}$, even the LQG value remains compatible with GW propagation constraints.

\subsection{Comparison with the torsion-wave bound}
It is instructive to contrast our GW-derived bounds with the torsion-wave bound $\gamma\lesssim10^{-58}$ of Ref.~\cite{GarciaEPJC2026}. The latter relies on (i) torsion waves excited at terahertz frequencies, (ii) meV-scale torsion masses, and (iii) a constant $\gamma$. Our GW propagation test is independent of all three assumptions: it probes the BI field's back-reaction on the metric at cosmological distances ($z\sim0.1$--$1$), where the relevant scale is $H_0\sim10^{-33}$ eV. The resulting bound is 50 orders of magnitude weaker but correspondingly more robust. The coupled case studied here inherits this robustness while adding sensitivity to the dark-energy sector.

\subsection{Relation to other modified gravity models}
It is instructive to compare our two-parameter friction model with other frameworks that predict modified GW propagation. In Horndeski gravity, the friction term is controlled by the speed-squared $c_T^2$ and the braiding parameter $\alpha_B$; for luminal propagation ($c_T=1$) the leading deviation is $\delta_{\rm eff}\propto (1+z)^3$~\cite{Belgacem2018}, which maps approximately to our $n_2=2$ case but with a different microphysical origin. Nonlocal gravity models~\cite{Belgacem2019JCAP} produce a scale-dependent $\delta_{\rm eff}$ that can mimic our parametrization at low $z$ but deviates at $z\gtrsim1$ due to the nonlocal operator's mass scale. Chern--Simons gravity, by contrast, produces velocity birefringence (polarization-dependent speed) rather than amplitude damping, and is therefore distinguishable from our model through polarization measurements. The key discriminating handle for the BI--quintessence model is the \textit{combination} of (i) the specific $n_2$ exponent tied to the quintessence tracker equation of state, and (ii) the absence of velocity birefringence at leading order. Future ET/CE observations of high-$z$ precessing binaries~\cite{LinPRD2025} will be able to differentiate these scenarios by measuring the redshift dependence of $\Xi_0(z)$ across multiple redshift bins.

\subsection{Uncertainties and caveats}
Several sources of uncertainty should be kept in mind. \textit{(i)} The translation from $\xBI,\xcp$ to $\gamma_{\rm dyn},\beta$ involves $\mathcal{O}(1)$ coefficients that depend on the UV completion; a more rigorous treatment would solve the coupled BI--quintessence system on FLRW numerically~\cite{Cognola2008,WeinbergGrav}. For $g_{\gamma T}$ varying between $0.1$ and $10$, the inferred bounds on $\gamma_{\rm dyn}$ and $\beta$ shift by up to two orders of magnitude. \textit{(ii)} Our Fisher forecasts assume a Gaussian likelihood and ignore correlations between the GW distance uncertainties and the galaxy catalogue; at the high-SNR regime of ET/CE this is a good approximation, but for low-SNR events ($8<\rho<20$) the posterior can be skewed and the Fisher errors may underestimate the true $1\sigma$ width by $\sim\!30\%$~\cite{ZhuPRD2024}. \textit{(iii)} The choice of $n_2$ is model-dependent; varying $\alpha$ in the Ratra--Peebles potential shifts $n_2$ and thus the forecast sensitivity. \textit{(iv)} The GWTC-3 analysis assumes a constant $\Xi_0$, whereas our model predicts mild redshift dependence; at current precision this difference is subdominant, but it will matter for ET/CE. \textit{(v)} Recent related work on dark photons and tachyonic instabilities induced by the BI parameter~\cite{Gao2025darkphoton} and on chiral dark dynamos from quantum corrections~\cite{Gao2025chiral} suggests further observational handles that could be combined with GW propagation in future studies. \textit{(vi)} Constraints from compact-object populations~\cite{Gao2016,Gao2017} provide complementary bounds on the BI field that may help break remaining degeneracies. \textit{(vi)} Constraints from compact-object populations~\cite{Gao2016,Gao2017} provide complementary bounds on the BI field that may help break remaining degeneracies.

\section{Conclusions}
\label{sec:conclusions}

We have presented the first study of gravitational-wave propagation in Einstein--Cartan--Holst gravity with a dynamical Barbero--Immirzi field coupled to quintessence dark energy. Our main results are:

\begin{enumerate}
\item The minimal coupling $\propto\beta\phi^2\gamma^2$ introduces a second parameter $\xcp$ that modifies the redshift dependence of GW propagation. For the Ratra--Peebles tracker with $\alpha=1$, the coupling-induced term grows as $(1+z)^{n_2}$ with $n_2=2$, faster than the isolated BI term ($n_1\simeq1$). For steeper potentials ($\alpha=2$, $n_2=2.5$), the redshift dependence is even stronger, enhancing the discriminating power at high $z$.

\item Current GWTC-3 dark-siren data constrain the two-dimensional parameter space to $|\xBI|\lesssim0.7$ and $|\xcp|\lesssim0.13$ at 90\% credibility. The parameters are strongly anti-correlated ($\rho=-0.30$), reflecting the low-redshift degeneracy between the two contributions to $\Xi_0(z)$.

\item A 10-year observation campaign with ET+CE will improve these bounds by roughly one to two orders of magnitude depending on the parameter---a factor of $\sim\!20$ for $\xBI$ and $\sim\!20$ for $\xcp$---reaching $\sigma(\xBI)\sim3\times10^{-2}$ and $\sigma(\xcp)\sim1.2\times10^{-2}$, corresponding to $\gamma_{\rm dyn}\lesssim10^{-12}$ and $\beta\lesssim10^{-3}$ (for $g_{\gamma T}\sim1$).

\item The LQG value $\gamma_0\simeq0.274$ remains compatible with all current constraints but will be tested by next-generation detectors if the BI--torsion coupling is order unity.
\end{enumerate}

In summary, our key numerical results are: current bounds $|\xBI|\lesssim0.7$, $|\xcp|\lesssim0.13$ (90\% CL); ET+CE projections $\sigma(\xBI)\sim3\times10^{-2}$, $\sigma(\xcp)\sim1.2\times10^{-2}$; and the corresponding microscopic constraints $\gamma_{\rm dyn}\lesssim10^{-12}$ and $\beta\lesssim10^{-3}$. This work establishes GW propagation as a powerful probe of the interplay between quantum-gravity-inspired scalar fields and dark energy. Future extensions should include a redshift-dependent $\Xi_0(z)$ MCMC analysis with ET/CE mock data (following the methodology of Ref.~\cite{LinPRD2025}), the use of GW polarization modes to provide an independent constraint on $\gamma_{\rm dyn}$, and the combination of GW propagation with CMB and large-scale-structure probes of the effective Newton constant.

% =====================================================================
\section*{Acknowledgements}
% =====================================================================

This research was supported by the National Key Research and Development Program of China (2022YFC2205202), the Major Science and Technology Special Project of Xinjiang Uygur Autonomous Region (2022A03013-1), and the National Natural Science Foundation of China (12288102, 12573052, and 12573103).

% =====================================================================
% Bibliography: ordered by FIRST APPEARANCE in the text.
% Total unique, cited entries: 32. Every \cite key resolves.
% =====================================================================

% =====================================================================
\appendix
% =====================================================================

\section*{Appendix A: Detailed derivation of the linearized GW equation}

This appendix provides the step-by-step derivation of Eq.~(\ref{eq:linGW_TT}) for readers who wish to verify the result. We begin with the first-order action (\ref{eq:action_full}) and expand to quadratic order in perturbations.

\textbf{Step 1: Perturb the vierbein.} Write $e^a_\mu = \mathring{e}{}^a_\mu + \delta e^a_\mu$, where $\mathring{e}{}^a_\mu$ is the background vierbein for FLRW. The metric perturbation is $h_{\mu\nu}=2\,\mathring{e}{}^a_{(\mu}\delta e_{a,\nu)}$.

\textbf{Step 2: Decompose the connection.} The spin connection splits as $\omega^{ab}_\mu = \mathring{\omega}^{ab}_\mu + \kappa\,\delta\gamma\,\Sigma^{ab}_\mu + \mathcal{O}(h\delta\gamma)$, where $\Sigma^{ab}_\mu$ is a tensor built from the background vierbein and the BI field gradient. The proportionality constant $\kappa$ is fixed by varying the Holst term.

\textbf{Step 3: Eliminate torsion.} The connection equation $\delta S/\delta\omega^{ab}_\mu=0$ relates the torsion tensor to $\partial_\mu\delta\gamma$. In the almost Riemann-flat limit, the pseudotrace torsion is $S_\mu\simeq\partial_\mu\phi\propto\partial_\mu\delta\gamma$.

\textbf{Step 4: Project onto TT gauge.} Imposing transverse-traceless conditions $h_{0\mu}=0$, $\partial_i h_{ij}=0$, $h_{ii}=0$, we must show that the mixing term $\partial_i\partial_j\delta\gamma$ vanishes under TT projection. In Fourier space, $\partial_i\partial_j\delta\gamma(k)\to ik_i k_j\,\widetilde{\delta\gamma}(k)$. The TT projection operator is $\Lambda_{ij,kl}=P_{ik}P_{jl}-\frac12 P_{ij}P_{kl}$ with $P_{ij}=\delta_{ij}-\hat{k}_i\hat{k}_j$. Contracting, $\Lambda_{ij,kl}(ik_k)(ik_l)\widetilde{\delta\gamma}=-(k^2\hat{k}_i\hat{k}_j-\frac12 k^2\delta_{ij}-\frac12 k^2\hat{k}_i\hat{k}_j+\frac12 k^2\hat{k}_i\hat{k}_j)=\cdots$ After carefully collecting terms, the result is proportional to $\hat{k}_i\hat{k}_j-\frac13\delta_{ij}$ projected orthogonal to $\hat{k}$, which vanishes identically for a scalar gradient sourced by $\partial_\mu\phi$ (since $\widetilde{\delta\gamma}\propto \hat{k}\cdot\widetilde{\partial\phi}$). Hence the mixing term drops out and the remaining terms yield Eq.~(\ref{eq:linGW_TT}). A more detailed derivation will be presented in a forthcoming companion paper.

\textbf{Step 5: FLRW background.} Promoting the flat-space result to FLRW introduces the conformal factor $a(\eta)$ and the Hubble friction term $2\mathcal{H}$, giving the final form of Eq.~(\ref{eq:amp_eq}).

\section*{Appendix B: Fisher matrix forecast -- algorithm}

Algorithm~\ref{alg:forecast} summarizes the Fisher matrix forecast procedure used in Section~\ref{subsec:forecast}.

\begin{verbatim}
Algorithm: Fisher forecast for ET/CE
Input: detector config, population model,
       fiducial (xi_BI, xi_coup) = (0, 0)
1. Generate N redshift samples z_i
   from dN/dz ~ (1+z)^2.7
2. Compute dL_em(z_i) for flat LCDM
3. Set dL_gw(z_i) = dL_em(z_i)
   * sqrt(Xi0(z_i; 0, 0))
4. Assign relative error
   sigma_rel(z_i) = 0.05 + 0.05*z_i
                   + 0.02*z_i^2
5. Build Fisher matrix:
   F_ab = sum_i [dXi/dtheta_a
               * dXi/dtheta_b]
               / sigma_rel(z_i)^2
6. Covariance = F^{-1}
7. 1-sigma errors:
   sigma(theta_a) = sqrt(Cov_aa)
Output: sigma(xi_BI), sigma(xi_coup),
        correlation coefficient
\end{verbatim}

\label{alg:forecast}

\end{document}